\documentclass[journal,10pt,draftclsnofoot,onecolumn]{IEEEtran}

\usepackage{color}
\usepackage{hyperref}
\usepackage{amsmath}
\usepackage{amssymb}
\usepackage{graphicx}

\makeatletter

\usepackage{color}
\usepackage[table]{xcolor}

\definecolor{azure}{RGB}{51, 102, 153}
\definecolor{light}{RGB}{224, 224, 224}
\definecolor{lightblue}{RGB}{204, 204, 255}

\makeatother

\IEEEoverridecommandlockouts 

\begin{document}

\title{Safe and Secure Wireless Power Transfer Networks: Challenges and Opportunities in RF-Based Systems}

\author{
\IEEEauthorblockN{Qingzhi Liu\IEEEauthorrefmark{1}\IEEEauthorrefmark{2}\IEEEauthorrefmark{3}, Kas{\i}m Sinan Y{\i}ld{\i}r{\i}m\IEEEauthorrefmark{1}\IEEEauthorrefmark{3}, Przemys{\l}aw Pawe{\l}czak\IEEEauthorrefmark{1}, and Martijn Warnier\IEEEauthorrefmark{2}}\\ 
\IEEEauthorblockA{\IEEEauthorrefmark{1}Department of Electrical Engineering, Mathematics and Computer Science\\ 
\IEEEauthorrefmark{2}Department of Technology, Policy and Management\\ Delft University of Technology, The Netherlands\\ Email: \{q.liu-1, k.s.yildirim, p.pawelczak, m.e.warnier\}@tudelft.nl}
\thanks{\IEEEauthorrefmark{3}These authors contributed equally to this work.}
\thanks{Supported by the Dutch Technology Foundation STW under contract 12491.}
}

\maketitle

\begin{abstract}

RF-based wireless power transfer networks (WPTNs) are deployed to transfer power to embedded devices over the air via radio frequency waves. Up until now, a considerable amount of effort has been devoted by researchers to design WPTNs that maximize several objectives such as harvested power, energy outage and charging delay. However, inherent security and safety issues are generally overlooked and these need to be solved if WPTNs are to be become widespread. This article focuses on safety and security problems related WPTNs and highlight their cruciality in terms of efficient and dependable operation of RF-based WPTNs. We provide a overview of new research opportunities in this emerging domain.

\end{abstract}

\section{Introduction}
\label{sec:Intro}

Over the past decade, powering cyber-physical systems has emerged as a growing challenge arising from increasing demand for mobility, sustainable operation and large-scale deployment. Conventional power cords lost their favor dramatically since they prohibit mobility and large-scale deployments. Batteries, on the other hand, increase weight, cost and ecological footprint of hardware, and their replenishment is generally impractical. Fortunately, recent advancements in technology have allowed the transfer of electromagnetic energy from a power source to receiver devices over the air, so called \emph{wireless power transfer} \emph{(WPT)}. In today's world, WPT is becoming more prevalent by means of considerable number of commercial products already on the market, e.g. WiTricity (\href{http://witricity.com/}{witricity.com}), uBeam (\href{http://ubeam.com/}{ubeam.com}), Ossia (\href{http://www.ossia.com/}{ossia.com}), Artemis (\href{http://www.artemis.com/}{artemis.com}), Energous (\href{http://www.energous.com/}{energous.com}), Proxi (\href{http://powerbyproxi.com/}{powerbyproxi.com}) and Powercast (\href{http://www.powercastco.com/}{powercastco.com}).

Thus far, WPT techniques has advanced towards two major directions. \emph{Non-radiative} techniques employ either inductive or magnetic resonant coupling for energy transfer to a wide range of appliances at short distances. In contrast, \emph{radiative} techniques use electric field of the electromagnetic waves, typically radio frequency (RF) waves, as an energy delivery medium to transfer power over long distances. RF-based WPT exhibits many other practical advantages such as provision of energy to many receivers simultaneously by means of its broadcast nature, low complexity, size and cost for the energy receiver hardware and suitability for mobile devices. Currently, RF-based WPT has already reached a sufficient level to charge low-power embedded devices like sensors and RFID tags in practice. Without any doubts, future advancements in energy transfer efficiency will enable it to be an indispensable component of numerous embedded systems~\cite{rf_powered_computing_gollakota_2014}, e.g. wireless cameras~\cite{wispcam_2015} or sensors.

\subsection{RF-Based Wireless Power Transfer Networks}

\begin{figure*}
\centering
\includegraphics[width=\columnwidth]{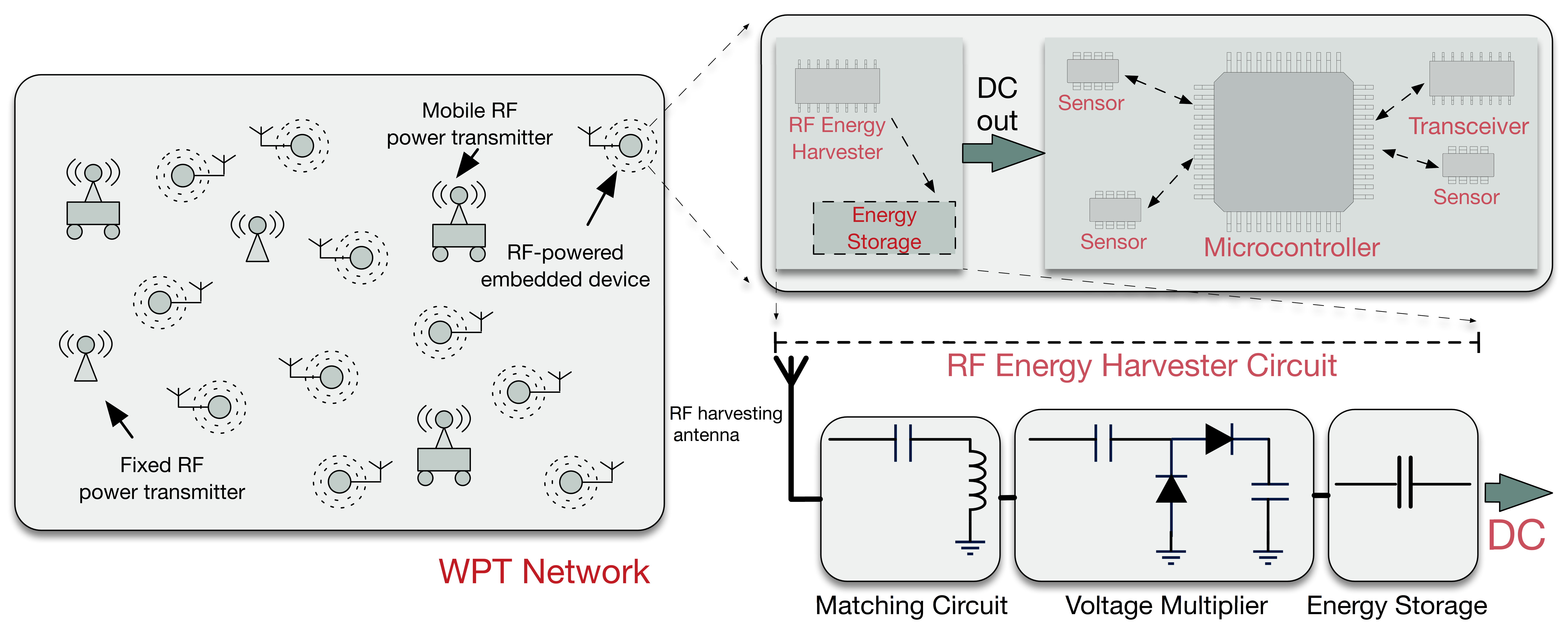}
\caption{\label{fig:wptn}A prospective WPTN architecture. Energy receivers such as computational RFIDs and wireless sensors are equipped with a RF energy harvester circuit that converts radio waves into DC power. The circuit is composed of an antenna (or an antenna array), a matching component, a voltage multiplier and an energy storage (capacitor or battery). The stored energy is used to power microcontroller, transceiver and sensors.}
\end{figure*}

As RF-based WPT technologies become more relevant in practice, the natural next step is the deployment of specialized networks of dedicated WPT nodes, namely \emph{wireless power transfer networks (WPTNs)}~\cite{liu15gwptn}, aimed at transferring sufficient power to nearby energy receiver (ER) devices over the air. Energy transmitter (ET) devices forming WPTNs are capable of controlling their transmit power and time/frequency of the waveforms in order to charge different types of ERs demanding different levels of energy. Each ER is equipped with a harvester circuit that converts the received RF signal to a DC signal to charge its built-in battery. A prospective WPTN architecture is depicted in Fig.~\ref{fig:wptn}.

Up until now, considerable amount of effort has been devoted by the researchers to design WPTNs maximizing several objectives such as harvested power~\cite{krikidis_tcom_2013}, energy outage~\cite{huang_arxiv_2012} and charging delay~\cite{fu_infocom_2013}. In contrast, inherent safety and security issues in WPTNs have not drawn attention by the research community yet. Concisely, traditional techniques dealing with secrecy of data and security of communication channel are not enough on their own to prevent security and safety holes during power transfer. Basically, wirelessly transmitted energy can be neither encrypted nor authenticated to ensure confidentiality and integrity which leads to power transfer channels being sensitive to outsider and insider attacks that might harm the health of its users as well as intercept its efficient transfer operation~\cite{secure_cooperative_kang_2015}. 

\subsection{Motivation}

\begin{figure*}
	\centering
	\includegraphics[width=\columnwidth]{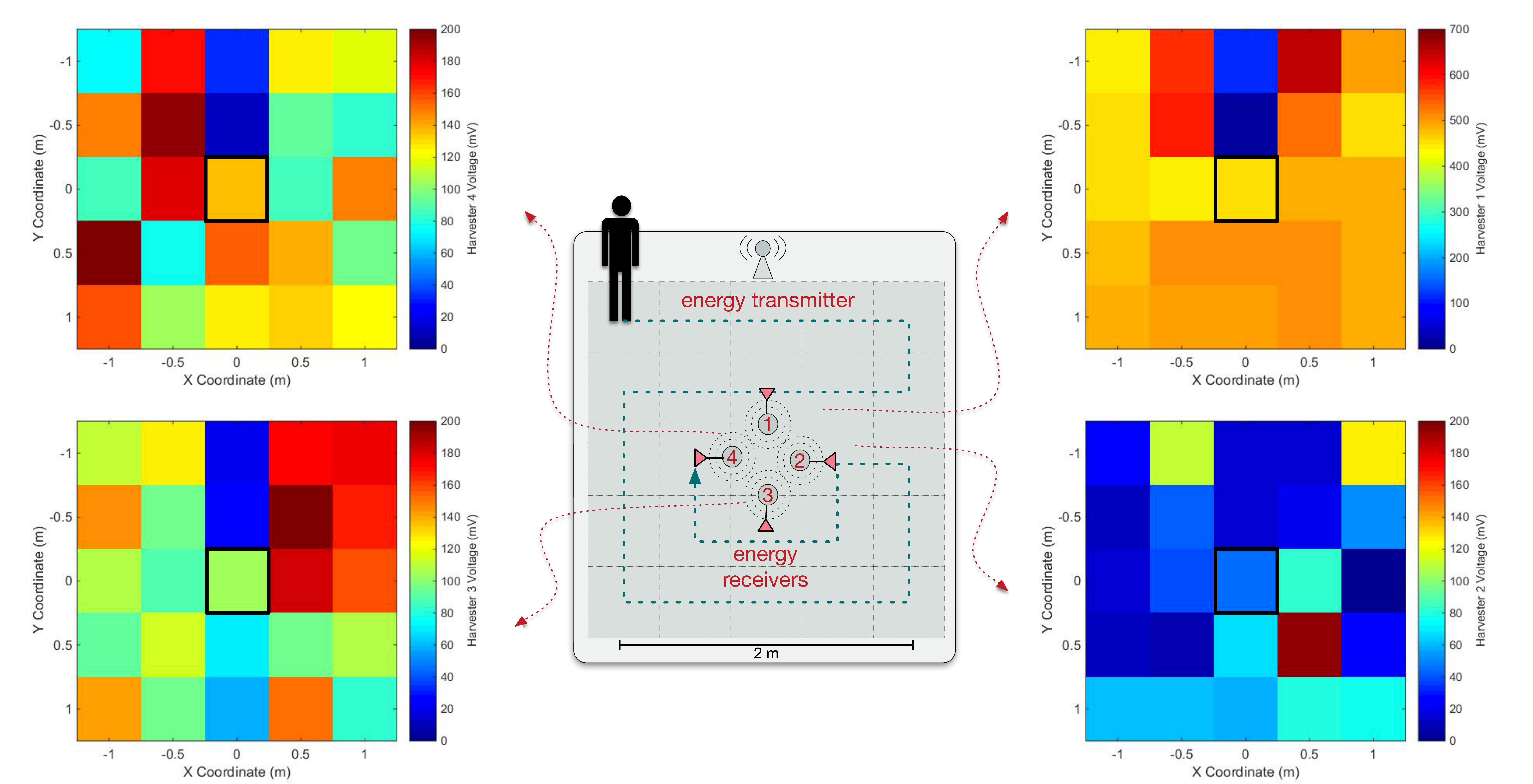}
	\caption{The measured voltage level at the receiver (ER) when a person is present at different locations of an university office. Each square in each of four heatmaps represents a measured voltage by the harvester (located at the center of a room) with a person located at the marked position. Center square (marked by a black line) represents a measurement with no person present.}
	\label{fig:harvested_voltage_finger_print_result}
\end{figure*}

For the sake of clarification, let us show how a small-scale WPTN formed by using off-the-shelf hardware can be abused to disclose the existence of nearby persons and thus violate anonymity. To this end, we deployed one Powercast TX91501 as ET and four P2110 evaluation boards with 9\,dBi patch antenna as ERs at fixed positions on a 2.5$\times$2.5\,meter area, as presented in Fig.~\ref{fig:harvested_voltage_finger_print_result}. In this setup, both ET and ERs had a ground clearance of 1 meter, each ER's antenna was directed towards a different wall, and ET was pointing towards ER (1). Conceptually, we divided the deployment area into square cells and assumed the midpoint of the cell at the center as the reference point (0,0) of our coordinate system. 

At the beginning of our experiment, we turned on the ET and measured the initial harvested voltage level for each ER when there was nobody in the deployment area. Thereafter, we collected voltage levels from each ER for each cell when a person was standing in the midpoint of the corresponding cell. Voltage was measured with an oscilloscope at the D$_{\text{OUT}}$ of ER by reading voltage after D$_{\text{OUT}}$ stabilized. We present our measurements using heat maps for each ER, where the middle cell of each heat map presents the initial harvested voltage levels and other cells present harvested voltage levels when humans are present.

We have the following observations. Firstly, the harvested energy is very sensitive to environmental conditions---even the existence of a person in the deployment area can change the harvested energy considerably. Secondly, any ER can monitor the existence of nearby persons to some extent by observing the change in its harvested voltage; it can even report its observations to an attacker without the consent of users. For instance, ER (1) in Fig. \ref{fig:harvested_voltage_finger_print_result} can easily detect if a person is standing in front of it, on the other hand ER (4) can detect if there is somebody on its left side. To conclude: the absence of security mechanisms might lead to unexpected vulnerabilities in WPTNs, as such in this example.   

\begin{figure}
	\centering
	\includegraphics[width=0.65\columnwidth]{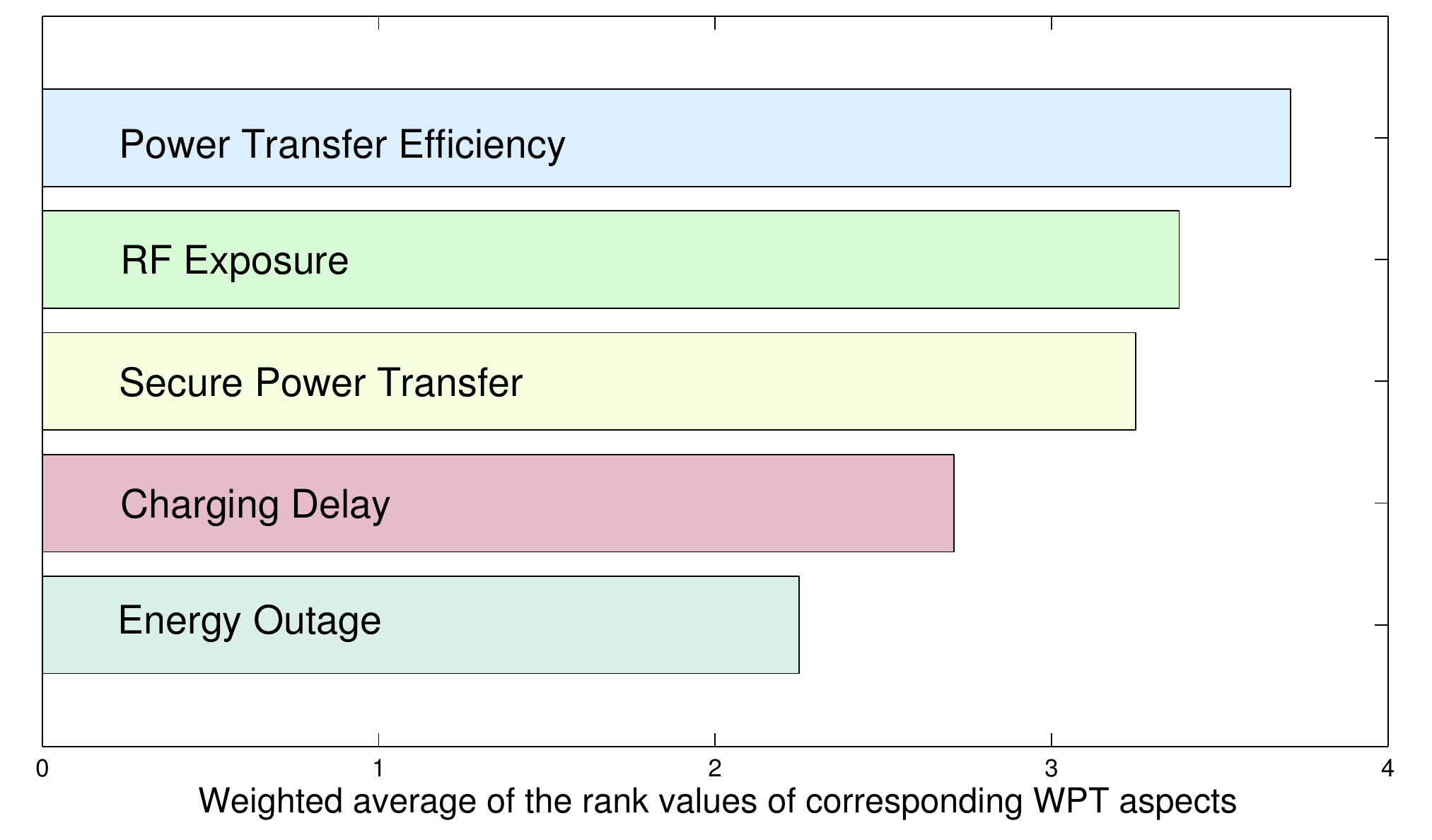}
	\caption{\label{fig:survey} We asked a selected group of 17 scientists (selected from Google Scholar list of `label:wireless\_power\_transfer' and from recently published papers in this domain' list is available upon request) working in WPT domain to rank the concepts of power transfer efficiency, charging delay, energy outage, RF exposure and secure power transfer according to their potential to become a barrier to widespread acceptance of WPTN. The researchers selected `1' for the least important and `5' for the most important according to their perception. The weighted average of the results indicate that safety and security aspects of WPT are promising points to be addressed in future WPT studies.}
\end{figure}

We believe that the aforementioned issues have great potential to become a barrier to widespread acceptance of WPTNs. To support our argument, we conducted a survey among top researchers working in the field, see Fig.~\ref{fig:survey} that indicates safety and security aspects are around the corner of being the main concern driving WTPN research. 

To the best of our knowledge this article is the first to introduce particular safety and security related problems that will lead to a roadmap for future research in WPTNs. First, we start by presenting underlying challenges for safe wireless power transmission in the sense of health impairments. Based on this background, we reveal particular security attacks that abuse several aspects in WPTNs, such as safety regulations and power transfer channel to intercept the operation of WPTNs. We point out several directions for the detection and elimination of the security attacks. Finally, we summarize and conclude the article.

\section{Challenges for Safe Wireless Power Transmission}
\label{sec:Safety}


To this end, several institutions established exposure limits for the radiation power, e.g. the International Commission on Non-Ionizing Radiation Protection (ICNIRP) in Europe allows a power density of 457.5 $\mu$W/cm$^{\text{2}}$ for the 915\,MHz frequency band used commonly for WPT (a detailed survey of the regulatory framework pertinent to wireless power transfer (WPT) systems is given in~\cite{kalialakis2014wpt}). Therefore, a safe-charging WPTN must comply with the RF exposure regulations and guarantee that none of its users are exposed to EMR that exceeds the safety threshold. 

\subsection{Safety Issues in Practice}
\label{subsec:safety issues}

In practice, environmental dynamics bring about additional difficulties to ensure EMR safety. First of all, the wireless power density is hard to estimate and control. For instance, the radio exposure can be over the limit due to reflection and refraction of the signals originated from other wireless devices. Second, ensuring safety where end-users are allowed to deploy new ETs and modify the locations of existing ETs/ERs at run-time is challenging. Several ETs can be active simultaneously, aimed at charging ERs collaboratively to satisfy energy demands of the receivers as fast as possible meanwhile optimizing the transferred energy. On the other hand, as more ETs are deployed, end users might be exposed to more radiation. Unfortunately, determining a power transfer schedule for ETs in order to maximize power transfer and ensure EMR safety is an NP-hard problem~\cite{dai_infocom_2014}. In such a dynamic system, one should guarantee the exposure safety considering run-time influence of unpredictable end-user actions. Third, for scenarios where ERs are wearable, body movements might lead to unpredictable exposure influence on different parts of the body.

Even though providing straightforward solutions to ensure power transfer safety might prevent RF exposure to some extent, several side effects might appear in practice. As an example, suppose that an ER is currently being charged by an ET. Let a neighboring ER with an almost-depleted battery sends a charging request to another ET. If that ET is turned on and starts transmitting energy, it might be the case that the RF exposure exceeds the safety threshold. However, if it remains turned-off, the neighboring ER might stop operating. Therefore, extra attention should be paid to such \emph{charging deadlocks} in order to detect and break them, which might be challenging in practical scenarios.

\subsection{Safe Power Transfer is not Easy}
\label{subsec:safety not easy}

Although wireless power density is the preferred quantity for safety regulations in millimeter wave frequencies, it does not rely on power absorption in tissues. On the other hand, temperature has a great potential to become a quantity for 
safety regulations in the future due to its higher relevance to hazardous damages, such as on eyes \cite{safe_generations_2015}. Therefore, practical attempts and regulations by now are not sufficient to realize safe power transfer and potential health impairments may strongly prevent wide-spread acceptance of WPTNs. We believe that extensive research efforts are required for the detection and prevention of unsafe power transfer operations in practice. On the other side, safety regulations can also be abused to create several security attacks, as we present in the following sections.

\begin{figure*}
\centering\includegraphics[width=0.75\columnwidth]{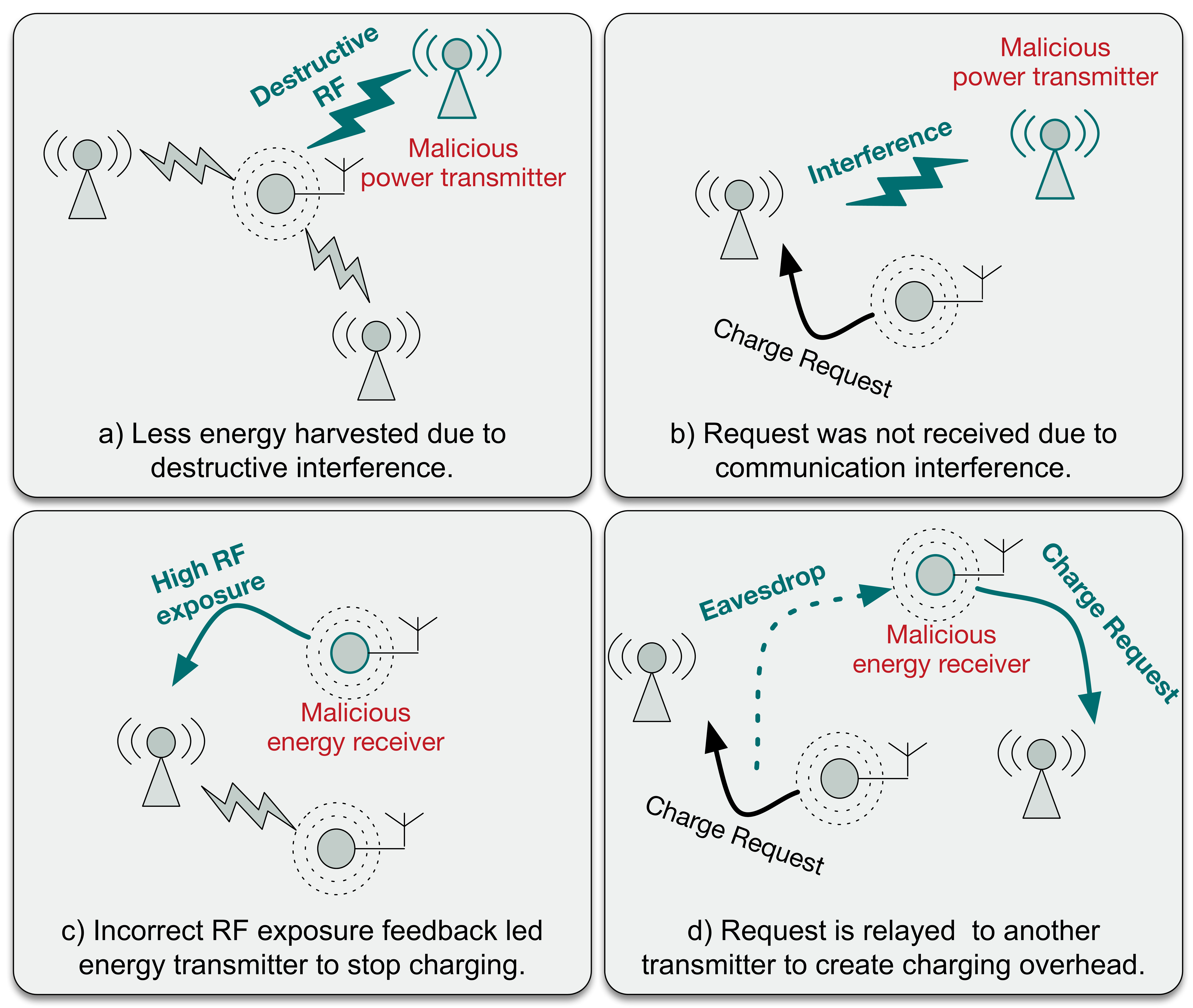}
\caption{\label{fig:dos}Schematic representation of different types of attacks in WPTNs: a) beamforming attack; b) jamming attack; c) charging attack; d) spoofing attack.}
\end{figure*}

\section{Security Attacks in Wireless Power Transfer Networks}

We start by outlining our assumptions about the underlying WPTN and security goals. Then we dive into particular attacks arising from the lack of security mechanisms. 

\subsection{Security Model and Goals}

It is assumed that ERs and ETs are equipped with transceivers to perform communication using different and inherently insecure wireless channels than that are used for power transfer. \emph{Malicious} nodes are defined as ERs and ETs that do not exhibit trustworthy behavior. The goal of the attacker is to degrade the power transfer efficiency and ruin the operation of the WPTN by introducing malicious nodes separately or capturing existing ones by programming them wirelessly. Malicious nodes can eavesdrop radio packets, duplicate/modify their contents, generate new ones to inject into the network and send incorrect information upon request, such as wrong battery or RF exposure levels. A secure WPTN should guarantee its integrity and availability, and preserve its sustainable, safe and efficient power transfer operation in the presence of malicious nodes. 

\subsection{Safety Attacks}

Safety regulations can be abused to degrade charging performance of ETs, even to force them stop working leading to a denial of service. In particular, safe charging can be ensured by controlling transmission parameters of ETs, such as power, frequency and transmission duration according to received feedback from ERs about RF exposure. However, if a malicious ER does not want its neighbors to be charged, it can always report that the RF exposure is over the safety limit. In such a case, ETs should either turn-off their transceivers or reduce their transmission power. Therefore, the more safety attacks, the less efficient ERs are charged and the shorter their operation time due to the possibly of depleted batteries. 

As emphasized previously, it is almost impossible to estimate wireless power density precisely due to its sensitivity against environmental dynamics. Better measurement and estimation techniques are required for ETs to obtain the radio power distribution without feedback from ERs. To this end, a solution would be the deployment of a dedicated sensor network to collect RF exposure values from the environment, eliminating the need of receiving feedback from ERs. However, this introduces additional overhead and cost to WPTNs.

\subsection{Charging Attacks}

ETs should receive requests and feedback from ERs to transfer them power in an efficient manner. However, malicious ERs may generate redundant or selfish requests and incorrect feedback deliberately to decrease the overall power transfer efficiency of WPTNs. 

\subsubsection{Freerider ERs}

ETs equipped with omni-directional antennas can be considered as public energy sources since any ER inside their coverage can harvest energy. To exemplify, when any ET receives a charging request from a specific ER and starts charging it, nearby ERs also harvest energy although they did not request it. While it might be desirable to charge multiple ERs simultaneously, thereby minimizing charging overhead, if this behavior is not controlled, then less ERs would prefer sending charging requests. Unfortunately, this might introduce overhead for ERs explicitly requesting energy. Even worse, ETs might not be able to optimize themselves since they will be unaware of the system-wide energy demands of ERs.

\subsubsection{Greedy and Cheating ERs}

Since ERs get the largest benefit when they always harvest energy, \emph{greedy} ERs might send charging requests to ETs continuously which may lead to other ERs receiving less power. This is even more problematic when ETs are equipped with directional antennas and beamform power only to the greedy ERs, which might prevent other ERs from receiving power. Therefore, ETs should implement intelligent and fair power transfer mechanisms. Since it is very challenging to estimate harvested energy precisely rather than measuring, ETs might prefer receiving feedback from ERs to collect information about their energy levels. Using this information, ETs can optimize their power transmission parameters. However, even though they have sufficient energy to perform their operations, \emph{cheating} ERs might report that their current energy level is low or that they harvested less energy in order to receive more power from ETs.

\subsubsection{How to Prevent Charging Attacks?}

For system-wide efficient and correct WPT operation, robust and resilient approaches are required to prevent freerider, greedy and cheating ER attacks. The natural first step is the detection of malicious nodes, which requires new methods for the estimation of harvested energy precisely, so that inaccurate information from greedy and cheating ERs can be crosschecked. One approach can be checking whether the received information is plausible and if incorrectness is likely, it can either be discarded or corrected using historic information. Moreover, rather than relying on a single ER, information received from multiple ERs can be considered to  identify malicious nodes. Another solution might be calibrating ETs to the reflected loss they will experience in their normal environment. For fixed ETs, a simple initialization routine that is done at the time of installation can determine how much loss is they should experience in the environment when it is ensured that no ERs are active nearby.  Combining this with beamforming could help the detection of unauthorized ERs. 

After detection, new techniques should be developed either to penalize malicious nodes or to reward secure ones to achieve fair charging. Regarding freeriders, ETs can modify their RF transmission parameters at run-time, e.g. frequency and power, and share them only with the ERs that requested energy, so that only they are charged efficiently and less power is transferred to freeriders. 

\subsection{Interference Attacks}

WPTNs cannot be isolated from their environment. Interference generated by some other devices using the same frequency band might lead to significant charging performance degradation. Interference might be generated deliberately by an attacker in order to intercept WPT operation. 

\subsubsection{Beamforming Attacks}

When multiple ETs emit RF waves at the same frequency band simultaneously, the received waves by ERs can either be constructive or destructive. In constructive interference, where the phase differences between the transmitted signals are negligible, the received power is greater than that of individual energy waves. On the contrary, destructive interference occurs when the phase difference is large, leading to less harvested energy than that of the individual energy waves~\cite{naderi_twc_2014}. Not surprisingly, destructive interference is a potential threat, which can be produced by an attacker deliberately to decrease or destroy harvested energy at ERs and intercept energy harvesting operation potentially leading to a full collapse of system functionality.

\subsubsection{Jamming Attacks}

Another type of attack is creating interference deliberately to intercept the communication between ETs and ERs. Malicious nodes can use one-channel jamming or swipe between all communication channels to jam them all. When the charging requests from ERs are blocked, ETs will not be informed about the energy demands and therefore might stop charging. This might result in ERs running out of energy and them to stop operating. 

\subsubsection{How to Prevent Interference Attacks?}

An important step to prevent interference attacks is interference detection, which can be achieved by turning off power transmission and communication periodically, listening in the network if there is a suspicious RF transmission and deciding if there is an attacker outside of the network. However, attackers can also synchronize themselves to the interference detection period so that they will remain silent, leading to mis-detection. Therefore, more robust and sophisticated methods are required in order to detect the existence of interference. After interference detection, ETs should dynamically adapt their transmission parameters so that ERs are not affected by the malicious nodes and they receive energy efficiently. It is an open question how such an adaption can be realized and a network-wide consensus is achieved for secure operation. Promising solutions for jamming attacks can be based on interference shaping techniques, such as interference alignment~\cite{elayach2013interferencealignment} that allows ETs to propagate signals that align at the ER so that each receiver is able to decode its intended signal and the aggregated interference effect is either minimized or eliminated. Although such techniques will allow ERs to receive data packets in the presence of jamming attacks, they are not easy to realize in practice and it is an open question how they can be implemented in WPTNs. 

\subsection{Spoofing Attacks}

In WPTNs, ERs and ETs broadcast beacons that contain various types of information, such as the energy level, device identity, etc. Malicious nodes can eavesdrop on the information of the other devices and use this information for their own benefits. For instance, if a malicious node can predict the time another ER will run out of energy, it can request energy just before that time it in order to prevent ETs from sending energy and to stop that node's operation. What is worse, a malicious node might receive and store a charging request beacon from a particular ER and impersonate it by rebroadcasting to another ET to increase its RF exposure, create interference or decrease throughput of the system. A standard solution for spoofing attacks can be the use of digital signatures that provide both data integrity and authentication to detect impersonating nodes. However, it is challenging to devise and implement such solutions for resource constrained devices~\cite{trappe2015lowenergysecurity}.

\begin{table*}
	\centering
	\caption{\label{tab:layers}Safety and Security Concerns in WPTNs and Their Potential Countermeasures}
	\begin{tabular}{|l|l|}
		\hline 
		\textbf{Type of security attack } & \textbf{Possible countermeasure(s)} \\ \hline\hline 
		Beamforming & Dynamic power and frequency adaptation\\ \hline 
		Jamming & Scheduling (e.g. FDMA, TDMA), interference alignment \\ \hline 
		Monitoring & Periodic scanning of the communication channel \\ \hline 
		Safety & RF-exposure and temperature estimation\\ \hline
		Charging & Harvested energy estimation\\ \hline
		Spoofing & Authorization, authentication\\ \hline
		Application & Signature check\\ \hline 
	\end{tabular}
\end{table*} 

\subsection{Application Attacks}

Another source of attacks is malicious applications running on ERs. A malicious application that consumes energy deliberately, might force an ER to send frequent charging requests to the ETs to intercept charging operation or decrease system efficiency. Such applications might also force ERs to stop operating by depleting their batteries. One solution is running trusted applications by checking application signatures. If a malicious application is found, it should be deactivated by ERs. Another approach might be energy-aware scheduling of applications so that every application has a pre-defined energy budget for its operations.

\subsection{Monitoring Attacks}
\label{Monitoring}

The attacks discussed so far can be considered as active attacks since they alter the operation of WPTNs explicitly. On the other hand, monitoring attacks are passive and it is challenging to detect them since they do not involve any alteration of the data flowing through WPTNs. To clarify, WPTNs can also be considered as wireless monitoring networks in the sense that malicious ER devices can also receive energy from ETs and they might disclose private information without the consent of WPTN users using the harvested energy. As an example, a malicious ER can be equipped with special sensors that collects measurements and notify an attacker about the potential events inside the WPTN.

A solution to detect monitoring attacks can be listening to the communication channel periodically in order to detect information exchange between malicious ERs and an attacker. However, if malicious communication is performed at a different channel, detection becomes more complicated. 

\subsection{A Summary of Attacks}

We summarize WPT security issues discussed so far in Table~\ref{tab:layers} by presenting each security attack and its potential countermeasure(s) that can be implemented to mitigate its effect. An important point worth mentioning is that countermeasures ensuring secure and safe energy transfer should be implemented as small and energy-efficient protocols. More precisely, each issue should be solved with low-power and limited resource requirements since wirelessly-powered embedded devices are resource constrained in general. This introduces additional difficulties since conventional safety and security solutions are resource hungry~\cite{trappe2015lowenergysecurity}. 

\section{Conclusions}

RF-based WPTNs have already become a promising solution to power a wide range of low-power embedded devices over the air. So far, a considerable amount of research has been devoted in this field. However, safety and security aspects are generally overlooked by the community. In this paper, we show that WPTNs are vulnerable to a variety of attacks. We present several research challenges and unanswered research questions in the area of safety and security that together provide a roadmap for future WPTN research. 

\bibliographystyle{IEEEtran}
\bibliography{IEEEabrv,references}

\end{document}